\documentstyle[aps,preprint,amsfonts]{revtex}
 \tightenlines
\begin{document}
\draft
\title{On the Avalanche-finiteness of Abelian Sandpiles}
\author{S. W. Chan$^1$ and H. F. Chau$^2$\footnote{To whom correspondence
 should be addressed.}}
\address{
 ~$^1$Department of Mathematics, Ohio State University, Columbus, OH 43210,
 U.S.A.\\
 ~$^2$School of Natural Sciences, Institute for Advanced Study, Olden Lane,\\
 Princeton, NJ 08540, U.S.A.
}
\date{\today}
\preprint{ISASNS-HEP-94/86}
\maketitle
\mediumtext
\begin{abstract}
 We prove a necessary and sufficient condition for an Abelian Sandpile Model
 (ASM) to be avalanche-finite, namely: all unstable states of the system can
 be brought back to stability in finite number of topplings. The method is also
 computationally feasible since it involves no greater than $\mbox{O} \left(
 N^3 \right)$ arithmetic computations where $N$ is the total number of sites of
 the system.
\end{abstract}
\medskip
\narrowtext
 The Abelian Sandpile Model (ASM), whose mathematical structure is first
 studied extensively by Dhar \cite{ASM}, is one of the few class of models of
 self-organized criticality in which a lot of interesting physical properties
 can be found analytically. The model consists of a finite number of sites
 labeled by an index set $I$. For each site $i\in I$, we assign an integer
 $h_i$ called the local height to it. Whenever the local height of a site
 exceeds a threshold (which is fixed to 0 for simplicity), the site is called
 unstable and it will transport some of its local heights (or sometimes called
 particles at that site) to the other sites in the coming timestep according to
\begin{equation}
 h_j \longrightarrow h_j - \Delta_{ij} \hspace{0.3in} \mbox{whenever~} h_i > 0
 \mbox{.}
\end{equation}
 $\Delta$ is called the toppling matrix whose elements satisfies
\begin{mathletters}
\begin{equation}
 \Delta_{ii} > 0 \hspace{0.25in} \forall\ i\in I \mbox{,} \label{E:Delta_Dia}
\end{equation}
 and
\begin{equation}
 \Delta_{ij} \leq 0 \hspace{0.25in} \forall\ i\ne j \mbox{.}
 \label{E:Delta_OffDia}
\end{equation}
 Sometimes, we may further require that particles cannot be created in the
 re-distribution process, which means that
\begin{equation}
 \sum_{j\in I} \Delta_{ij} \geq 0 \hspace{0.25in} \forall\ i\in I \mbox{.}
 \label{E:Delta_Conserve}
\end{equation}
\end{mathletters}
\par\noindent
 Toppling is repeated until all sites become stable again. The whole process of
 toppling is collectively known as an avalanche. The system is driven by adding
 a unit amount of particles onto the sites randomly and uniformly after the
 system regains stability.
\par
 It is, therefore, important to know the condition(s) for which the system
 relaxes to a stable state in a finite number of topplings. In fact, almost all
 previous studies of the Abelian Sandpile Model assume that such a relaxation
 can be taken in a finite number of steps. An Abelian Sandpile is said to be
 {\bf avalanche-finite} if and only if every unstable system configuration can
 regain stability in a finite number of steps via toppling. In this paper, we
 prove a necessary and sufficient condition for avalanche-finiteness. Actually
 our proof works equally well on integer or real toppling matrices.
\par
 A system configuration, stable or not, can be regarded as a point in the space
 ${\Bbb R}^N$ where $N$ is the total number of sites in the system. Moreover,
 toppling of particles can be regarded as a translation in ${\Bbb R}^N$. If
 site $i$ of a system configuration $\alpha$ is unstable, then the system
 configuration will become $\alpha - \Delta_i$ in the next timestep where
 $\Delta_i$ is the $i$-th row of the toppling matrix $\Delta$
 \cite{GASM,FigureCut}.
\par
 Dhar showed that the order of toppling does not affect the outcome of the
 final stable state in an avalanche. In this respect, the model is commutative.
 He then went on to prove that the average number of toppling occurs in site
 $j$ given that a particle is introduced in site $i$, $G_{ij}$, is given by
\begin{equation}
 G_{ij} = \left( \Delta^{-1} \right)_{ij}
\end{equation}
 provided that $\Delta$ is invertible \cite{ASM}.
\par
 Now we are going to prove a necessary and sufficient condition for
 avalanche-finiteness.
\par \bigskip
\noindent {\it Proposition~1:} An Abelian Sandpile Model with toppling matrix
 $\Delta$ is avalanche-finite if and only if there exist $n_i \in {\Bbb Z}^{+}
 ~\forall i\in I$ such that $\sum_i n_i \Delta_{ij} > 0$ for all $j\in I$.
\par \medskip
\noindent {\it Proof:} ($\Rightarrow$) Consider the unstable configuration
 $\alpha = (1,1,\ldots ,1)$. Avalanche-finiteness implies the existence of
 $n_i \in {\Bbb Z}^{+}$ such that $\alpha - \sum_i n_i \Delta_i$ is a stable
 configuration. As a result, $1 - \sum_i n_i \Delta_{ij} \leq 0$ for all $j\in
 I$ and hence it is proved.
\par
 ($\Leftarrow$) Suppose $\Delta$ is not avalanche-finite, then there exists an
 $\alpha\in {\Bbb R}^N$ which cannot regain stability by toppling in a finite
 number of steps. Let $J$ be the set of all unstable sites at this moment.
 Define $\beta^{(1)} = \left( \beta^{(1)}_i \right)_{i\in I}$ by
\begin{equation}
 \beta_j^{(1)} = \left\{ \begin{array}{ll} 1 & \hspace{0.15in} \mbox{if~} j\in
 J \\ 0 & \hspace{0.15in} \mbox{if~} j\not\in J \end{array} \right. \mbox{.}
\end{equation}
 Then in the next timestep, $\alpha$ will become $\alpha - \beta \Delta$. As
 the toppling will never stop, we will get an infinite sequence of vectors
 $\{ \beta^{(i)} \}$ by repeating the above process. Now if 1 occurs finitely
 many times in $\{ \beta^{(i)}_j \}_{i\in {\Bbb Z}^{+}}$, then site $j$ will be
 stable eventually. Let $I'$ be the set of all sites which become stable in a
 finite number of steps. Clearly $I \neq I'$ or else the toppling stops in
 finite number of steps. It is easy to see that $\Delta_{ij} = 0$ for $i\in
 I\backslash I'$ and $j\in I'$. Besides, Eq.~(\ref{E:Delta_OffDia}) together
 with $\sum_{i\in I} n_i \Delta_{ij} > 0$ implies that $\sum_{i\in I\backslash
 I'} n_i \Delta_{ij} > 0$ for all $j\in I\backslash I'$. But this means the
 local height for sites in $I\backslash I'$ will be negative after sufficiently
 long time, which is impossible.
\hfill $\Box$
\par \bigskip
\noindent {\it Proposition~2:} An Abelian Sandpile Model with toppling matrix
 $\Delta$ is avalanche-finite if and only if $\det\Delta \neq 0$ and $G_{ij}
 \equiv \left( \Delta^{-1} \right)_{ij} \geq 0$ for all $i,j \in I$.
\par \medskip
\noindent {\it Proof:} Let $P_i = \{ \sum_{j\neq i} b_j \Delta_j : b_j \geq 0
 ~\forall j \}$ be the non-negative span of $\{ \Delta_j \}_{j \in I \backslash
 \{ i \} }$, and $C = \{ \sum_{j\in I} b_j \Delta_j : b_j \geq 0 ~\forall j \}$
 be the cone generated by $\{ \Delta_j \}_{j\in I}$. Eq.~(\ref{E:Delta_OffDia})
 tells us that $\bigcup_{i\in I} P_i \cap \left( {\Bbb R}^{+} \right)^N =
 \emptyset$.
\par
 We consider the situation where $\det\Delta \neq 0$ first. As $\{ \Delta_j
 \}_{j\in I}$ is linearly independent, $\bigcup_{i\in I} P_i$, the boundary of
 $C$, disconnects ${\Bbb R}^N$. So either (i) $C \supset \left( {\Bbb R}^{+}
 \right)^N$; or (ii) $C \cap \left( {\Bbb R}^{+} \right)^N = \emptyset$.
\par
 Case(i): $~\Leftrightarrow \left( {\Bbb R}^{+} \right)^N \Delta^{-1} \subset
 \left( {\Bbb R}^{+} \cup \{ 0 \} \right)^{N}$. That is, $\left( {\Bbb R}^{+}
 \right)^N \Delta^{-1}$ is an open cone in $\left( {\Bbb R}^{+} \cup \{ 0 \}
 \right)^{N}$. Therefore, $\left( {\Bbb R}^{+} \right)^N \Delta^{-1}$ contains
 (infinitely many) positive integer lattice points. Hence result follows from
 Proposition~1.
\par
 Now we consider the remaining possibility where $\det\Delta = 0$. As $\{
 \Delta_j \}_{j\in I}$ is linearly dependent, one can check that $\bigcup_{i\in
 I} P_i = C$, hence $C \cap \left( {\Bbb R}^{+} \right)^{N} = \emptyset$. Thus,
 for any $n_i \in {\Bbb Z}^{+} ~\forall i\in I$, there exists $j\in I$ such
 that $\sum_i n_i \Delta_{ij} \leq 0$. By Proposition~1, the system is not
 avalanche-finite.
\hfill $\Box$
\par \bigskip
 With the above two propositions at hand, we are ready to prove our main
 theorem.
\par \bigskip
\noindent {\it Theorem~1:} An Abelian Sandpile Model with toppling matrix
 $\Delta$ satisfying Eqs.~(\ref{E:Delta_Dia}) and~(\ref{E:Delta_OffDia})
 is avalanche-finite if and only if $\det\Delta > 0$ and $G_{ij} \equiv \left(
 \Delta^{-1} \right)_{ij} \geq 0$ for all $i,j\in I$.
\par \medskip
\noindent {\it Proof:} By Proposition~2, it suffices to show that ``the ASM is
 avalanche-finite implies $\det\Delta > 0$''. We prove this claim by induction
 on $N$.
\par
 Obviously, our theorem holds when $N = 1$. Suppose it is true for $N = k-1$.
 Now for $N = k$, Proposition~2 assures that $\det\Delta \neq 0$. Let $\Delta'$
 be the $(k-1)\times (k-1)$ sub-matrix formed by deleting the $k$-th row and
 $k$-th column of $\Delta$. Using Eq.~(\ref{E:Delta_OffDia}) and Proposition~1,
 $\Delta'$ is avalanche-finite with $k-1$ sites. So by induction hypothesis,
 $C_{kk} (\Delta ) = \det\Delta' > 0$ where $C_{kk} (\Delta )$ denotes the
 $(k,k)$-th cofactor of $\Delta$. By Proposition~2, $(\det\Delta ) \cdot C_{kk}
 (\Delta ) \geq 0$. Therefore, $\det\Delta > 0$. So it is proved.
\hfill $\Box$
\par \medskip
\noindent {\it Remark~1:} For $N = 1,2,3$, one can easily check by direct
 computation that an Abelian Sandpile Model is avalanche-finite if and only if
 $\det\Delta > 0$.
\par \medskip
\noindent {\it Remark~2:} For $N \geq 4$, $\det\Delta > 0$ alone is not a
 sufficient condition for avalanche-finiteness. Consider the system with
 $\Delta$ given by
\begin{equation}
 \Delta = \left[ \begin{array}{rrrr} 1 & -2 & 0 & 0 \\ -1 & 1 & 0 & 0 \\ 0 &
 0 & 1 & -2 \\ 0 & 0 & -1 & 1 \end{array} \right] \oplus I_{N-4}
\end{equation}
 with $\alpha = (1,0,0,\ldots ,0)$. Direct computation verifies that the system
 is not avalanche-finite.
\par \bigskip
 Finally, if we restrict ourselves to the case where particles cannot be
 created during the toppling process, a simple criterion for
 avalanche-finiteness is found. But before we state our finding, we first
 report two important lemmas.
\par \bigskip
\noindent {\it Lemma 1:} $\det\Delta \geq 0$ if $\Delta$ satisfies
 Eqs.~(\ref{E:Delta_Dia})--(\ref{E:Delta_Conserve}).
\par \medskip
\noindent {\it Pf:} Consider the elementary row operations of
\begin{equation}
 [ \mbox{row~} i ] \longrightarrow [ \mbox{row~} i ] - \frac{\Delta_{i1}}{
 \Delta_{11}} [ \mbox{row~} 1 ] \mbox{,}
\end{equation}
 for $i\geq 2$, with resulting matrix $\Delta^{(1)} = \left[
 \Delta^{(1)}_{ij} \right]$. Now $\Delta^{(1)}_{i1} = 0$ for $i\geq 2$,
 $\Delta^{(1)}_{ij} \leq 0$ for $i\neq j$, $i\geq 2$. Consider
\begin{equation}
 \Delta^{(1)}_{22} = \Delta_{22} - \frac{|\Delta_{21}|\ |\Delta_{12}|}{
 \Delta_{11}} \geq \Delta_{22} - |\Delta_{21}| = \Delta_{22} + \Delta_{21}
 \geq 0 \label{E:Equality}
\end{equation}
 with equality holds if and only if $\Delta_{11} = |\Delta_{12}|$ and
 $\Delta_{22} = |\Delta_{21}|$. Eqs.~(\ref{E:Delta_Dia})
 and~(\ref{E:Delta_OffDia}) implies that $\Delta_{1j} = \Delta_{2j} = 0$ for
 $j\geq 2$. So $\Delta^{(1)}_{22} = 0$ if and only if $\Delta^{(1)}_{2} =
 \vec{0}$. Then $\det\Delta = 0$.
\par
 On the other hand, if $\Delta_{22}^{(1)} > 0$, then $\Delta^{(1)}_{ii} > 0$,
 $\Delta^{(1)}_{ij} \leq 0$ for $i\neq j,~\geq 2$. Also, it is easy to verify
 that $\sum_{j > 1} \Delta^{(1)}_{ij} \geq 0$. Hence we can repeat our
 argument on a smaller sub-matrix until either (i) we get to the point where
 the equality holds in Eq.~(\ref{E:Equality}) and hence $\det\Delta = 0$; or
 (ii) we eventually end up with an upper triangular matrix with positive
 diagonal elements, and hence $\det\Delta > 0$.
\hfill $\Box$
\par \bigskip
\noindent {\it Lemma~2:} $C_{ij} (\Delta ) \geq 0$ for all $i,j\in I$ if
 $\Delta$ satisfies Eqs.~(\ref{E:Delta_Dia})--(\ref{E:Delta_Conserve}). Here
 $C_{ij} (\Delta )$ denotes the $(i,j)$-th cofactor of $\Delta$.
\par \medskip
\noindent {\it Proof:} Note that $C_{ij} (\Delta ) = \det\Xi$, where $\Xi$ is
 the $N\times N$ matrix with elements given by
\begin{equation}
 \Xi_{pq} = \left\{ \begin{array}{ll} 1 & \hspace{0.15in} \mbox{if~} p = i
 \mbox{~and~} q = j \\ 0 & \hspace{0.15in} \mbox{if~} p = i \mbox{~and~} q \neq
 j \\ 0 & \hspace{0.15in} \mbox{if~} p \neq i \mbox{~and~} q = j \\
 \Delta_{pq} & \hspace{0.15in} \mbox{otherwise} \end{array} \right. \mbox{.}
\end{equation}
 It is easy to check that the row sums of $\Xi$ are all non-negative except for
 the $j$-th row. Let $k$ be an integer greater than $\sum_{r\in I\backslash \{
 j \}} |\Delta_{jr}|$, and consider the row operation $[ \mbox{row~} j]
 \longrightarrow [ \mbox{row~} j] + k [\mbox{row~} i]$. Then it is obvious
 that the row sums of this new matrix are all non-negative and hence by
 Lemma~1, $C_{ij} (\Delta ) = \det\Xi \geq 0$.
\hfill $\Box$
\par \medskip
\noindent {\it Remark~4:} Thus if we further require $\Delta$ to be invertible,
 then $G_{ij} \equiv \left( \Delta^{-1} \right)_{ij} \geq 0$ for all $i,j \in
 I$.
\par \bigskip
\noindent {\it Corollary~1:} An Abelian Sandpile Model with toppling matrix
 $\Delta$ satisfying Eqs.~(\ref{E:Delta_Dia})--(\ref{E:Delta_Conserve}) is
 avalanche-finite if and only if $\det\Delta > 0$.
\par \medskip
\noindent {\it Proof:} Direct application of Theorem~1 and Lemma~2.
\hfill $\Box$
\par \bigskip
 In conclusion, we found a necessary and sufficient condition for
 avalanche-finiteness for Abelian Sandpiles. In the event that particles may
 be created during the toppling process, avalanche-finiteness can be examined
 by looking at the determinant and the elements of the inverse of the toppling
 matrix $\Delta$, a process requiring not greater than $\mbox{O} \left( N^3
 \right)$ arithmetic computations. Thus, our method is computationally
 practical. Furthermore, our finding is consistent with the interpretation
 that $\left( \Delta^{-1} \right)_{ij}$ is the average number of topplings
 occurs in site $j$ given a particle is introduced in site $i$.
\par
 If we further restrict ourselves to the case where particle cannot be created
 in toppling, our avalanche-finiteness testing scheme can be further
 simplified to a simple test in the sign of $\det\Delta$, which again can be
 done efficiently in the computational point of view.
\acknowledgments{This work is supported in part by DOE grant
 DE-FG02-90ER40542.}

\end{document}